# Unveiling cosmic X-ray/gamma-ray sources from their scattering screen

Jonathan E. Grindlay
Harvard-Smithsonian Center for Astrophysics, Cambridge, MA 02138, USA
Email: josh@head.cfa.harvard.edu

Accretion-powered cosmic X-ray sources are rarely clean. The very nature of their energy source – matter liberating its gravitational free-fall energy by accreting onto a compact object (e.g. black hole or neutron star) -- implies the radiation produced must traverse, at least partially, the matter providing (upstream) the source of accretion power. The electrons in the accreting matter will scatter the radiation produced by their preceding flow with the cross section given by the Thomson value ($\sigma_T = 6.4 \times 10^{-25}$ cm$^2$) for scattering of photons with energy $\varepsilon$ on electrons of temperature $T_e$ when both $\varepsilon$ and $kT_e$ are $<< = m_e c^2 = 511$keV, the electron rest mass energy. Otherwise the full energy dependent Compton cross section in the Klein-Nishina regime applies (for which $\sigma_C \sim \sigma_T(1 - 2x + ....)$, for $x = \varepsilon/m_e c^2 <<1$, or $\sigma_C = (3\sigma_T/8x)(\ln 2x + 1/2)$ for $x >>1$), but in either case the resulting spectrum of accretion-powered radiation is *Comptonized* with physical conditions in the accretion flow (or surrounding clouds) then imprinted on the initial spectrum. The scattering optical depth $\tau$ and electron temperature $T_e$ both alter the emergent photon spectrum, for which average photon energies are increased by factors $\Delta\varepsilon/\varepsilon \sim 4kT_e/m_e c^2$ if $kT_e >> \varepsilon$ or are decreased if $kT_e << \varepsilon$. The seminal paper, entitled "Comptonization of X-rays in Plasma Clouds. Typical Radiation Spectra" (Sunyaev and Titarchuk 1980; hereafter ST80), provided the first complete description of these processes with application to four distinct, but ultimately related, problems of interest for understanding cosmic X-ray source spectra and variability. The paper has had a strong impact on the development of high energy astrophysics where the understanding of compact objects and energetic processes (e.g., near black holes) requires that the several sources of radiation, all unresolvable spatially, are understood and deconvolved spectrally and temporally.

Apart from its mathematical elegance and physical completeness, the beauty of this paper is that it enabled the scattering screen to not only be lifted but in fact used for new physical understanding of accretion sources arising from the combination of relatively cool and optically thick accretion disks with their Comptonizing coronae, or extended and generally optically thin hotter atmospheres. The principal use of the Comptonization models for fitting X-ray spectra is to explain the "power law" or hard radiation components in X-ray spectra that appear (in some spectral states) to be added on to the otherwise relatively soft spectra that are expected from accretion disk models. Whereas the X-ray spectrum of an optically thick and geometrically thin accretion disk was first derived in the landmark paper of Shakura and Sunyaev (1973), the importance of Compton scattering in an overlying disk atmosphere had been considered for the early observed spectra and variability of the black hole candidate Cyg X-1. Shapiro, Lightman and Eardley 1976) studied this by obtaining an approximate analytic solution of the Kompaneets equation (Kompaneets 1956); however, the focus was on the region nearest the BH and not the overall structure of the source. ST80 provided a full treatment, and opened new realms of study of the physically more complete nature and structure of compact objects, generally, as revealed by their X-ray (as well as hard X-ray/soft $\gamma$-ray) spectra and variability.

Beginning in the 1980s with the broad-band (~0.5-20 keV), non-imaging X-ray satellites *Tenma* and then *Ginga*, which provided spectra and timing studies of bright galactic X-ray binaries, and continuing in the 1990s with the higher sensitivity, broader-band, and moderate spatial resolution imaging satellite *BeppoSAX*, the application of Comptonization models to galactic X-ray binaries containing accreting neutron stars and black holes and also moderately bright active galactic nuclei (AGN) was carried out by many investigators. The initial Comptonization model of ST80 became known as `compST` and was incorporated into the standard X-ray spectral analysis software package XSPEC (Arnaud 1996). This was extended with `compTT` (Titarchuk 1994) to include relativistic effects and to work well in both the optically thin and thick regimes. A further generalization to include Comptonization from the bulk motion of the accretion flow itself was the BMC model of Titarchuk, Mastichiadis and Kylafis (1997), also then widely used in XSPEC. However, this did not include the electron recoil effect and so diverged at low energies. This has been recently fixed, and incorporated in XSPEC, with an empirical model, SIMPL, that is particularly useful for determining the inner disk radius of the seed photon distribution that is Comptonized into an observed spectrum (Steiner, Narayan, and McClintock 2008). This is especially useful for constraining BH spin, which is constrained by the innermost stable circular orbit (ISCO) of matter at the inner edge of the accretion disk around the BH. Another recent extension of the BMC model is the `compTB` model of Farinelli et al (2008), which includes a geometry-dependent factor and for which good spectral fits and source constraints were derived for six bright LMXBs containing neutron star primaries.

Comptonization models have primarily been applied to both X-ray binaries and AGN for which the broad spectral energy distribution (SED) is thermal or at least contains thermal components and is not dominated by obviously non-thermal (synchrotron, inverse Compton) processes. Most radio-quiet AGN are included, and it is the standard prescription for fitting Seyfert galaxy (low luminosity AGN) spectra, including Fe K emission line and reflection components from the disk, such as for NGC4151 (Zdziarski et al 2002). Thus it is surprising that the flat spectrum radio quasar 4C04.42 shows evidence of bulk Comptonization in addition to its obviously non-thermal broad band components, as discussed by de Rosa et al (2008). Bulk Comptonization has been suggested for explaining spectral flattening in other blazars as well (e.g. BGB1428+4217; Celotti et al 2007). Although this combined thermal/non-thermal picture is not yet required, it is entirely plausible that cold "blobs" could be entrained in the bulk flows of (some) jets, so further observations are needed.

Most applications of Comptonization and extensions from ST80 consider spectral distortions (**"Problems 2 – 4"** discussed by ST80) and do not consider "Problem 1" discussed by ST80: the temporal broadening of a flare due to the additional time for photon diffusion out of a scattering cloud. The temporal broadening case (Problem 1) was introduced to explain the temporal structure and spectral hardening noted in the discovery of an X-ray burst from the first identified X-ray burst source (Grindlay et al 1976), but has not been much considered as part of the Comptonization paradigm since. A new application, with a very different type of high energy burst source, may present itself: the fact that the two highest redshift Gamma-ray bursts (GRBs) both have source-frame durations ~1 s and are thus consistent with being Short GRBs (Kouveliotou et al 1993), yet their extreme luminosity and other factors point to them being Long GRBs. The recent record-redshift burst, GRB090423 (z ~8.2), and the second most distant, GRB080913 (z = 6.7), both have durations $T90_{ref}$ ~1sec in their source reference frame (Zhang and Zhang 2009). This may be an observational selection effect caused by the preferential detection of the most distant GRBs with shortest duration and thus peak flux. It may also stem from the preferential detection of GRBs with the largest bulk Lorentz factors $\Gamma$ in their jets, and thus the smallest beaming angles $\theta$ ~$1/\Gamma$ and therefore the smallest burst durations $T90_{ref} \sim R/(2c\Gamma^2)$ for emission from a source (colliding shocks) of radius R in the relativistic jet. However, it might also come from reduced Comptonization of the prompt emission by the external medium if these two highest-z GRBs are in significantly reduced metallicity hosts with little or no self-enrichment. A nearly pure H, He surrounding interstellar medium (with metallicity significantly below even PopII values, as expected at these high redshifts in the early Universe) would reduce the photon-scattering optical depth $\tau_o$ in the circum-burst environment and ISM (from which the afterglow arises) and thus the expected (ST80) pulse-broadening contribution, $\Delta T \sim 0.3\tau_o^2$, to the intrinsic $T90_{ref}$ value. This may be required if more high-z GRBs are found to show similar $T90_{ref}$ values, since low-z GRBs (e.g. GRB080319B) with the highestest inferred $\Gamma$ values and smallest jet opening angles (Racusin et al 2008) do not seem to have similarly low $T90_{ref}$ ~1sec values. Thus Comptonization effects may affect the appearance of even the most extreme events, GRBs, by again imprinting their signature on the observed radiation.